# Evaluating Wikipedia as a source of information for disease understanding

Eduardo P. García del Valle, Gerardo Lagunes García, Lucia Prieto Santamaría, Massimiliano Zanin
Centro de Tecnología Biomédica
Universidad Politécnica de Madrid
Pozuelo de Alarcón, Spain
{ep.garcia, lucia.prieto.santamaria}@alumnos.upm.es,
{gerardo.lagunes, massimiliano.zanin}@ctb.upm.es

Alejandro Rodríguez-González[1], Ernestina Menasalvas Ruiz
Centro de Tecnología Biomédica, ETS Ingenieros Informáticos
Universidad Politécnica de Madrid
Pozuelo de Alarcón, Spain
{alejandro.rg, ernestina.menasalvas}@upm.es

*Abstract*—The increasing availability of biological data is improving our understanding of diseases and providing new insight into their underlying relationships. Thanks to the improvements on both text mining techniques and computational capacity, the combination of biological data with semantic information obtained from medical publications has proven to be a very promising path. However, the limitations in the access to these data and their lack of structure pose challenges to this approach. In this document we propose the use of Wikipedia - the free online encyclopedia - as a source of accessible textual information for disease understanding research. To check its validity, we compare its performance in the determination of relationships between diseases with that of PubMed, one of the most consulted data sources of medical texts. The obtained results suggest that the information extracted from Wikipedia is as relevant as that obtained from PubMed abstracts (i.e. the free access portion of its articles), although further research is proposed to verify its reliability for medical studies.

## I. Introduction

The study of diseases as non-isolated elements and the understanding of how they resemble and relate to each other is crucial to provide novel insights into diagnostic decision, as well as in the identification of new targets and applications for drugs [1]. The complete sequencing of the human genome at the beginning of the 21st century represented a revolution in the study of the relationships between diseases. In combination with the growing availability of transcriptomic, proteomic, and metabolomic data sources, it should help to improve the classification of diseases. However, the use of these sources was affected by problems such as their fragmentation, heterogeneity, availability and different conceptualization of their data, which could affect the accuracy of the studies based on them [2]. The exploitation of the emerging sources of biological data in combination with existing textual data, such as clinical histories or scientific articles, allows researchers to compensate for these limitations. This is especially noticeable in genetic research, where having up-to-date information on complex processes involving genes, proteins and phenotypes is crucial [3].

The use of clinical records in disease research is recurrent in many studies, especially in those focused on symptoms. In 2007, Rzhetsky et al. [4] used the disease history of 1.5 million patients at the Columbia University Medical Center to infer the comorbidity links between disorders and proving that phenotypes form a highly connected network of strong pairwise correlation. In 2009, Hidalgo et al. [5] built a Phenotypic Disease Network (PDN) summarizing the connections of more than 10 thousand diseases obtained from pairwise comorbidity correlations reconstructed from over 30 million records from Medicare patients. Despite this demonstrated potential in pathological analysis, the access and use of clinical records in medical research is limited by several issues, including the heterogeneity of sources [6], ethical and legal restrictions and the disparity of regulations between countries [7].

The analysis of medical literature (i.e. books, journals, publications) has been used as an alternative to medical records. The increasing availability of retrieval engines such as PubMed or UKPMC, maintained by the US National Center for Biotechnology Information (NCBI) and the European Bioinformatics Institute (EBI), respectively, has boosted this approach [3]. Numerous studies cover the task of detecting medical concepts among other purely literary terms, when these sources are used. XueZhong et al. used large-scale medical bibliographic records from PubMed to generate a symptom-based network of human diseases, where the link weight between two diseases quantifies the similarity of their respective symptoms [8]. Okumura et al. [9] performed an analysis of the mapping between clinical vocabularies and findings in medical literature using OMIM as a knowledge source and MetaMap as the NLP tool. Following this idea, Rodríguez et al. [10] used web scraping and a combination of NLP techniques to extract diagnostic clinical findings from MedlinePlus articles about infectious diseases using MetaMap tool. In a further study, the same team compared the performance of MetaMap and cTakes in the same task [11].

In view of these studies, the use of medical texts is proving to be one of the most promising ways to infer disease-symptom relationships and unveil unknown connections

---

between them. However, this approach presents serious limitations that hinder its implementation. Probably the most important is the access to these texts. Although the ethical and legal limitations for clinical records are not applicable to medical publications, access to articles is often restricted and requires the payment of a subscription or accepting conditions that may prevent their exploitation [12]. For instance, in the particular case of PubMed, often only the abstract of the article can be accessed free of charge. Although in some cases open versions of the articles are available at PubMed Central (ncbi.nlm.nih.gov/pmc), they do not have a common structure, which complicates their mining.

As a consequence of these limitations, there is a need to explore new medical textual sources with free access. In the present work we propose the use of articles from Wikipedia - the online free encyclopedia Wikipedia - in the extraction of disease data that can be used in subsequent scientific and academic research. Wikipedia articles are contributed by users from all around the world and reviewed by volunteers and Wikipedia staff [13]. This often raises doubts about their reliability as an academic source, although there are studies that show that its accuracy is similar to other sources with higher levels of control [14] [15]. In order to evaluate its validity in the context of disease similarity research, we carried out an experiment to calculate the similarity between diseases based on the symptoms obtained from articles in Wikipedia and compared the results with those obtained using terms extracted from PubMed articles.

## II. METHOD AND MATERIALS

### A. Extraction, transformation and loading

First, we queried DBPedia - a project that extracts structured content (in RDF – Resource Description Framework) from the information in Wikipedia - to collect the links to all the Wikipedia articles categorized as *diseases*. Afterwards with the help of a web crawler, we visited the 8,161 obtained links and extracted those articles containing any reference to phenotypic manifestations. At the end of this mining phase, we had recovered 3,911 documents, containing 31,095 sections with relevant text and an average length of 438.68 characters. In a next phase, following the procedure described in [11], the MetaMap tool was used to extract a total of 9,937 UMLS concepts from all disease articles. The tool was configured to recover the semantic terms of type *sosy*, *diap*, *dsyn*, *fndg*, *lbpr* and *lbtr*, corresponding to concepts useful for the diagnosis process. After running the Term Validation Process (TVP), also described in [11], the list of concepts was reduced to 1,565 unique valid medical phenotypes associated to 3,595 diseases.

### B. Evaluation

As demonstrated in previous studies [8] [16], the similarity of the symptoms associated with two diseases can be used to measure the relationship between them. That is, the smaller the distance between the symptoms of two diseases, the greater their relationship is. To assess the quality of Wikipedia as a relevant source of information on diseases, we checked whether this condition was met using the extracted terms. In other words, we tried to verify whether the similarity of the extracted terms is higher, in a statistically significantly way, among related diseases (alternative hypothesis) than among unrelated diseases (null hypothesis). The rationale for this hypothesis, as well as a discussion of its possible limitations, will further be explored in Section IV.

First, to obtain a set of related diseases, we resorted to use a well-established disease classification system as proposed in previous studies [19] [20]. In particular, we obtained groups of similar diseases with a common second level class in the Disease Ontology (DO). For this purpose, we used different disease identifiers (e.g. OMIM, MeSH, ICD) to map the diseases extracted from Wikipedia with those in the DO and then, based on the *is_a* attribute of each disease, we looked up in the hierarchy to find its ancestor two levels below the root (i.e. *DOID:4 ! disease*). For each of the 86 second level classes containing diseases, we obtained all possible disease-disease pairs, resulting in a total of 21,760 relations. In order to compute the similarity of the diseases in each pair, the diseases were expressed as a vector of symptom concepts. For our experiment, we chose the following similarity measures, which are the main ones used in the literature [17]:

- Cosine similarity, computed as:

$$sim_{Cosine}(A,B) = \frac{|A \cap B|}{|A|^{\frac{1}{2}} + |B|^{\frac{1}{2}}} = \frac{\sum_{i=1}^{N} A_i B_i}{\sqrt{\sum_{i=1}^{N} A_i^2} \sqrt{\sum_{i=1}^{N} B_i^2}}$$

- Jaccard similarity, defined as:

$$sim_{Jaccard}(A,B) = \frac{|A \cap B|}{|A \cup B|} = \frac{\sum_{i=1}^{N} \min(A_i B_i)}{\sum_{i=1}^{N} \max(A_i B_i)}$$

- Dice similarity, defined as:

$$sim_{Dice}(A,B) = \frac{2|A \cap B|}{|A| + |B|} = \frac{2 \sum_{i=1}^{N} \min(A_i B_i)}{\sum_{i=1}^{N} (A_i B_i)}$$

where $A_i$ and $B_i$ *are the set of diagnosis concepts for respectively the diseases A and B*. Figure 1 shows an example of the two concept vectors corresponding to a disease pair in the same DO class.

| Disease vector A (presbyopia) | | |
|---|---|---|
| *i* | *Concept* | *UMLS CUI* |
| 1 | Asthenopia | C0004095 |
| 2 | Astigmatism | C0004106 |
| 3 | Blurred vision | C0344232 |
| 4 | Myopia | C0027092 |
| 5 | Retinoscopy | C0455973 |
| 6 | Visual impairment | C3665347 |

Cosine Similarity = 0.707
Jaccard Similarity = 0.5
Dice Similarity = 0.667

| Disease vector B (hyperopia) | | |
|---|---|---|
| *i* | *Concept* | *UMLS CUI* |
| 1 | Amblyopia | C0002418 |
| 2 | Asthenopia | C0004095 |
| 3 | Astigmatism | C0004106 |
| 4 | Blurred vision | C0344232 |
| 5 | Diabetes mellitus, NOS | C0011849 |
| 6 | Diplopia | C0012569 |
| 7 | Headache | C0018681 |
| 8 | Hyperopia | C0020490 |
| 9 | Myopia | C0027092 |
| 10 | Retinoscopy | C0455973 |
| 11 | Strabismus | C0038379 |
| 12 | Visual impairment | C3665347 |

Figure 1. Example of two diseases of the same DO class represented as vectors of their associated UMLS diagnosis concepts, and the similarities yielded by the three considered measures.

To ensure a meaningful comparison between diseases, only vectors with a minimum of 5 concepts were considered. This value was chosen based on the exploratory analysis of the datasets, described in the Results section.

Finally, we performed 100 paired t-test with the similarity values obtained for a set of 100 randomly selected related disease pairs and 100 randomly selected unrelated disease pairs. The p-values obtained for each similarity measure were adjusted with the Benjamini-Hochberg correction [18] and tested against a significance value α = 0.01.

### C. Comparison with PubMed

The previously described analysis was aimed at demonstrating the feasibility of using Wikipedia as a source for information on diseases. In a complementary way, and to evaluate its performance with respect to other sources of medical data, we also compared the results obtained with the terms extracted from Wikipedia with those obtained with corresponding to PubMed. It is worthy to mention that PubMed comprises more than 28 million citations for biomedical literature from Medline, life science journals, and online books, ant it has been proven to be a relevant source of information to calculate disease similarities [21] [22].

PubMed articles are annotated with descriptors from the Medical Subject Heading (MeSH) thesaurus. To obtain a set of PubMed articles related to each of the diseases extracted from Wikipedia, we made use of the search web service of the National Center for Biotechnology Information (NCBI), filtering by the MeSH descriptor of the disease. To reduce the execution time, we used the *[majr:noexp]* options to search a MeSH heading as a major topic and turn off the automatic explode (i.e. to prevent PubMed from searching more specific terms beneath that heading in the MeSH hierarchy). We also limited the results to a maximum of 100 documents per disease. In total, for the 1,407 diseases obtained from Wikipedia containing any identifier that could be mapped to MeSH through the DO, we retrieved data from 138,459 PubMed articles, including their title, their abstract and their URL, among others. The average abstract length is 669.58 characters. As in the case of Wikipedia articles, the MetaMap tool was then used to obtain the UMLS diagnosis concepts associated with each disease from the abstracts of their PubMed articles, to be later validated using the TVP validation. Altogether, 1,192 unique valid terms were obtained. Finally, by repeating the null hypothesis test previously described, we obtained the p-values for the PubMed data and compared them with those of the Wikipedia data. Outcome is detailed in the Results section.

If the information obtained from Wikipedia allowed to characterize the diseases with a precision similar to that obtained from PubMed, one would expect the terms obtained for the same disease to be very similar. To test this hypothesis, we calculated the similarity between pairs of the same disease described with terms obtained from the two sources and then compared it with that obtained with pairs of different diseases, following the same methodology as in the previous analysis.

The obtained values are detailed in the Results section. Figure 5 summarizes the described methodology.

### D. Materials

A Java application was developed for the data extraction and the similarity calculation. All data were collected in February 2018 and the files are available online [2].

## III. RESULTS

### A. Exploratory analysis of the obtained data

The average number of diagnosis concepts extracted per disease with data from up to 100 related abstracts in PubMed is 9.66, while the value is 12.29 for Wikipedia. This suggests that Wikipedia provides a greater amount of relevant data, taking into account that the volume of analyzed text is less than that of PubMed. Figure 2 shows the distribution of the number of concepts per disease for each source. While for PubMed probability distribution resembles a normal one, in the case of Wikipedia it follows a power-law, with a small number of diseases contain a large number of concepts. In both sources, 300 diseases contain less than 5 associated concepts, representing around the 20 percent of the total.

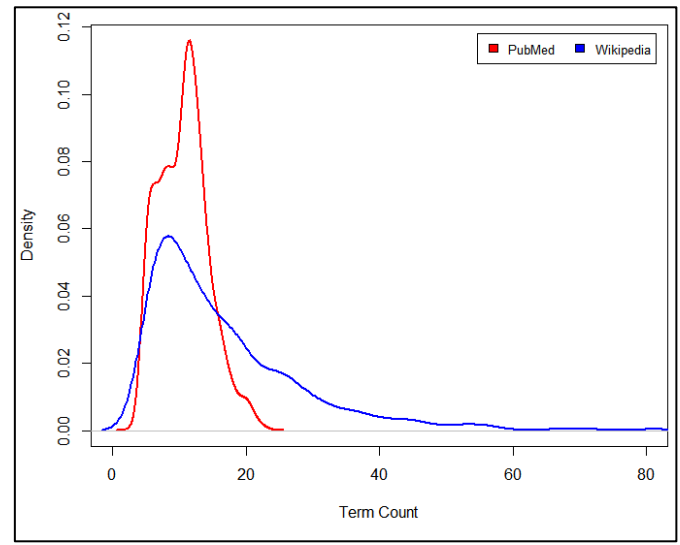

Figure 2. Distribution of validated concepts extracted per disease from the different sources.

In the case of Wikipedia, the disease with the highest number of detected concepts is *sarcoidosis*, with 83 terms. For this disease, only 13 concepts were extracted from PubMed abstracts. While some of them are common to both sources, like *skin lesion* or *complete atrioventricular block*, others like *exhaustion* or *redness of eye* only exist in PubMed data. On the other hand, the disease with the most concepts mined from PubMed is *cardiomyopathy*, with 23 terms compared to 13 in Wikipedia. Again, some concepts such as *heart failure* are common, but others like *edema* or *septicemia* are not. This suggests that the difference between the two sources lies not only in the number of concepts found per disease, but also in the concepts themselves.

---

[2] https://github.com/pantapps/cmbs2018

Table I reports the average concept count for the top-level disease classes in the DO. As observed, a comparable number of symptom concepts were retrieved for all main classes, meaning that the information available in both sources is not particularly biased towards any of them.

TABLE I. AVERAGE CONCEPT COUNT FOR THE TOP-LEVEL DISEASE CLASSES IN THE DISEASE ONTOLOGY, BY SOURCE.

| DO Class | Average concept count by disease | |
|---|---|---|
| | *Wikipedia* | *PubMed* |
| Disease by infectious agent | 14.936 | 9.12 |
| Disease of anatomical entity | 11.197 | 10.787 |
| Disease of cellular proliferation | 14.374 | 7.899 |
| Disease of mental health | 12.5 | 8.877 |
| Disease of metabolism | 9.403 | 10.201 |
| Genetic disease | 8.649 | 9.254 |
| Physical disorder | 10.522 | 8.857 |
| Syndrome | 14.936 | 12.087 |

*B. Same disease class*

Figure 3 represents the distribution of the similarities between terms of diseases of the same class, computed with the three considered measures. If the source is considered, the distributions for Wikipedia and PubMed are similar. However, the results vary considerably depending on the measure. As expected, the lowest similarities were obtained with Jaccard, since this coefficient is affected by the difference in the vector lengths. In other words, if vector *A* is much longer than vector *B*, even if *A* contains most terms in *B* ($A \cap B$ is large), as the total number of unique terms ($A \cup B$) is large in proportion, their Jaccard similarity is low. In contrast, computed similarities were higher when using Dice, and the highest with Cosine.

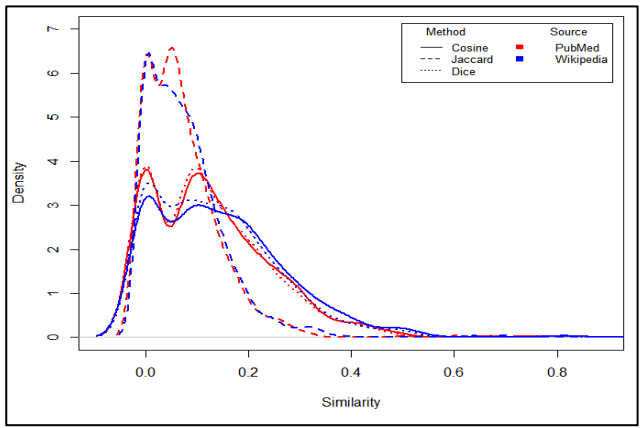

Figure 3. Similarity distribution for diseases of the same class.

A manual review of the results reveals that the higher similarities correspond to analogous diseases. For instance, using terms from Wikipedia we found that *adrenocortical carcinoma* is identical to *adrenal cortex cancer* (computed similarity was 1 for all measures) and *Guillain-Barre syndrome* is quite similar (a value of 0.894 with Cosine) to *Miller Fisher syndrome*, which belongs to the DO second level class *hypersensitivity reaction disease*. When we considered terms extracted from PubMed, *Brill-Zinsser disease* is close to *typhus* (Cosine similarity of 0.7), as expected according to their DO classification. For both sources, similarities are in general close to zero in disease pairs from different classes.

Table II contains the *p*-values obtained when evaluating the similarity between terms of diseases in the same class compared to the random case. Results are separated by data source (Wikipedia and PubMed) and similarity measure. In all cases, a significant effect (albeit just below the significance threshold) can be observed, suggesting that both Wikipedia and PubMed are an accurate source for disease classification [8] [16].

TABLE II. *P*-VALUES FOR DISEASES OF THE SAME CLASS

| Source | *P*-Values for Disease Similarity | | |
|---|---|---|---|
| | *Cosine* | *Jaccard* | *Dice* |
| *Wikipedia* | 0.0097 | 0.0098 | 0.0090 |
| *PubMed* | 0.0082 | 0.0076 | 0.0098 |

*C. Same disease*

Figure 4 depicts the distribution of the similarities between terms of the same disease in the different sources, computed with the three considered measures. Again, the obtained values were higher with Cosine and Dice coefficients than with Jaccard.

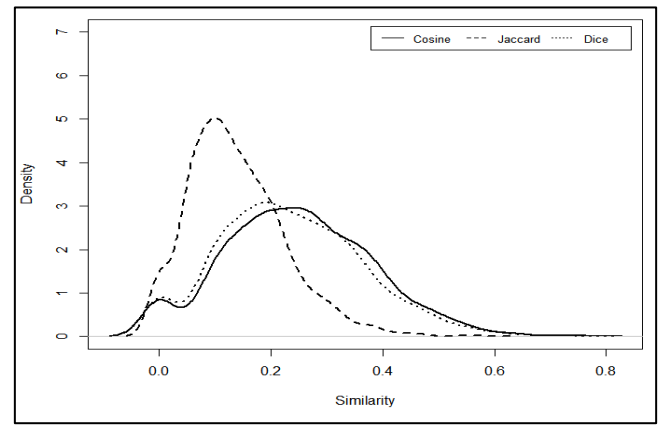

Figure 4. Distribution of computed similarities between terms of the same disease from different sources.

Table III shows the *p*-values obtained when evaluating the similarity between terms extracted from Wikipedia and PubMed for the same diseases, compared to that obtained in the random case. As we expected, values are well below the significance threshold, indicating that the information obtained from both sources is significantly coinciding. However, this does not imply that the data are identical, as denoted by the concept count distribution and similarity distribution previously described.

TABLE III. *P*-VALUES FOR THE SAME DISEASE IN DIFFERENT SOURCES

| Cosine | Jaccard | Dice |
| --- | --- | --- |
| 2.107E-18 | 6.633E-16 | 7.317E-18 |

## IV. DISCUSSION

The obtained results indicate that the data extracted from Wikipedia allow to calculate the similarity between diseases with a precision similar to what obtained using PubMed. While it is true that the complete PubMed corpus has not here been exploited, Wikipedia articles have shown to have a higher density of useful information (that is, diagnosis concepts) for the study of diseases based on their phenotypic manifestations. This may be due to the fact that, in the case of PubMed, we only used terms extracted from abstracts, in which the information about the diagnosis concepts of the disease is more limited than in Wikipedia articles.

The long-tailed distribution of the number of concepts by disease in Wikipedia possibly stems from the fact that certain diseases have been more extensively documented than others. Although this is also observed in PubMed, the distribution in this case is more bounded. This can be explained by a greater availability of documents per disease than in Wikipedia. It is finally worth noting that the statistical analyses here presented leverage on the hypothesis that similar diseases (i.e. diseases classified within the same DO) should share similar symptoms. This is supported by the idea that similar underlying mechanisms should yield similar manifestations; and has further been used in many scientific studies [8] [16]. Some voices of concern have nevertheless been raised, as a single name is sometimes used to define conditions with similar symptoms, yet raising from different mechanisms. This is the case, for instance, of primary progressing and relapsing-remitting multiple sclerosis [23]. The opposite is also true, namely that different conditions may be the result of the same molecular mechanisms - see for instance the onset of neuropathological lesions in Alzheimer's and Down's syndrome [24]. The statistical study of lists of symptoms, as the one here presented, may help shedding light on this topic.

## V. CONCLUSIONS

Despite its demonstrated potential for the analysis of the relationships between diseases and their symptoms/diagnosis elements, the exploitation of medical literature is hindered by factors such as its limited access and heterogeneity. In this document we proposed the use of Wikipedia as a source of structured and free-access text data. We evaluated its usefulness in the detection of relations between diseases based on its symptoms/diagnosis elements, and then compared its performance with that of PubMed. The obtained results show that Wikipedia can be as relevant a source as PubMed for this type of analysis.

## VI. STUDY LIMITATIONS AND FUTURE WORK

To limit the computation time of our experiment, the number of papers obtained from PubMed for each disease was limited to 100. The repetition of the study without this limit is proposed as future work, including a more exhaustive validation of the recovered data. Moreover, to obtain the disease-disease similarities, we used three well-known similarity measures. The application of alternative techniques for estimating such similarity, like the ones described by Okumura [25], is also proposed as future work.

Furthermore, it has to be noted that both Wikipedia and PubMed are living sources, i.e. they are constantly corrected and expanded; one of the advantages of the proposed methodology is that it can be executed periodically to track the variation in their performance. For example, the temporal evolution of the density of validated terms per analyzed document could be measured to see if it improves or worsens over time for each source.

ACKNOWLEDGMENT

This paper is supported by European Union's Horizon 2020 research and innovation programme under grant agreement No. 727658, project IASIS (Integration and analysis of heterogeneous big data for precision medicine and suggested treatments for different types of patients).

REFERENCES

[1] Kwang-I, Goh, Michael E. Cusick, David Valle, Barton Childs, Marc Vidal and Albert-László Barabási, "The human disease network", PNAS 2007 May, 104 (21) 8685-8690.

[2] Joseph Loscalzo, Isaac Kohane, Albert‐Laszlo Barabasi, "Human disease classification in the postgenomic era: A complex systems approach to human pathobiology", Published online 10.07.2007 in Molecular Systems Biology (2007) 3, 124.

[3] Dietrich Rebholz-Schuhmann, Anika Oellrich & Robert Hoehndorf, "Text-mining solutions for biomedical research: enabling integrative biology", in Nature Reviews Genetics volume 13, pages 829–839 (2012).

[4] Andrey Rzhetsky, David Wajngurt, Naeun Park and Tian Zheng, "Probing genetic overlap among complex human phenotypes", in PNAS 2007 July, 104 (28) 11694-11699.

[5] Hidalgo CA, Blumm N, Barabási A-L, Christakis NA, "A Dynamic Network Approach for the Study of Human Phenotypes", PLoS Comput Biol 2009 5(4): e1000353.

[6] Mike Conway, Richard L. Berg, David Carrell, Joshua C. Denny, Abel N. Kho, Iftikhar J. Kullo, et al., "Analyzing the Heterogeneity and Complexity of Electronic Health Record Oriented Phenotyping Algorithms", in AMIA Annu Symp Proc. 2011; 2011: 274–283. Published online 2011 Oct 22.

[7] Yip C, Han NLR, Sng BL, "Legal and ethical issues in research". Indian J Anaesth 2016;60:684-8.

[8] XueZhong Zhou, Jörg Menche, Albert-László Barabási, Amitabh Sharma, "Human symptoms–disease network", in Nature Communications volume 5, Article number: 4212 (2014).

[9] Okumura, T., Aramaki, E., Tateisi, Y., "Clinical vocabulary and clinical finding concepts in medical literature", in Proceedings of the International Joint Conference on Natural Language Processing Workshop on Natural Language Processing for Medical and Healthcare Fields, pp. 7–13 (2013).

[10] Rodríguez-González A, Costumero R, Martínez-Romero M, Wilkinson MD, Menasalvas-Ruiz E., "Diagnostic knowledge

extraction from MedlinePlus: an application for infectious diseases", in 9th International Conference on Practical Applications of Computational Biology & Bioinformatics (PACBB 2015), 2015, June 3-5, Salamanca, Spain.
[11] Rodríguez-González Rodríguez González, Alejandro; Costumero Moreno, Roberto; Martínez Romero, Marcos; Wilkinson, Mark Denis y Menasalvas Ruiz, Ernestina, "Extracting diagnostic knowledge from MedLine Plus: a comparison between MetaMap and cTAKES Approaches", in Current Bioinformatics, v. 375 (2015) ; pp. 1-7. ISSN 1574-8936.
[12] Elisabetta Poltronieri, Elena Bravo, Tiziana Camerini, Maurizio Ferri, Roberto Rizzo, Renata Solimini and Gaetana Cognetti, "Where on earth to publish? A sample survey comparing traditional and open access publishing in the oncological field", in Journal of Experimental & Clinical Cancer Research 2013 32:4.
[13] Andrew Feldstein, "Deconstructing Wikipedia: Collaborative Content Creation in an Open Process Platform", in Procedia Social and Behavioral Sciences 26 (2011) 76 – 84.
[14] Giles, J. "Internet encyclopaedias go head to head: Jimmy Wales' Wikipedia comes close to Britannica in terms of the accuracy of its science entries". 2005 Nature. 438 (7070): 900–1.
[15] Thompson, Neil and Hanley, Douglas, "Science Is Shaped by Wikipedia: Evidence From a Randomized Control Trial", in MIT Sloan Research Paper (February 13, 2018). No. 5238-17.
[16] Sachin Mathur, Deendayal Dinakarpandian, "Finding disease similarity based on implicit semantic similarity", in Journal of Biomedical Informatics. Volume 45, Issue 2, April 2012, Pages 363-371.
[17] N. Ljubesic, D. Boras, N. Bakaric and J. Njavro, "Comparing measures of semantic similarity," ITI 2008 - 30th International Conference on Information Technology Interfaces, Dubrovnik, 2008, pp. 675-682.
[18] Benjamini, Yoav; Hochberg, Yosef (1995). "Controlling the false discovery rate: a practical and powerful approach to multiple testing" (PDF). Journal of the Royal Statistical Society, Series B. 57 (1): 289–300. MR 1325392.
[19] Žitnik M, Janjić V, Larminie C, Zupan B, Pržulj N. "Discovering disease-disease associations by fusing systems-level molecular data". Scientific Reports. 2013;3:3202. doi:10.1038/srep03202.
[20] Sun K, Gonçalves JP, Larminie C, Pržulj N. "Predicting disease associations via biological network analysis". BMC Bioinformatics. 2014;15(1):304. doi:10.1186/1471-2105-15-304.
[21] Li Y, Agarwal P. "A pathway-based view of human diseases and disease relationships". Plos One 2009; 4: e4346, doi: 10.1371/journal.pone.0004346
[22] Hoehndorf R, Schofield PN, Gkoutos GV. "Analysis of the human diseasome using phenotype similarity between common, genetic, and infectious diseases". Scientific Reports. 2015;5:10888. doi:10.1038/srep10888
[23] Polman, C. H., Reingold, S. C., Banwell, B., Clanet, M., Cohen, J. A., Filippi, M., Lublin, F. D., "Diagnostic criteria for multiple sclerosis: 2010 revisions to the McDonald criteria". Annals of neurology (2011), 69(2), 292-302.
[24] Selkoe, D. J. "The molecular pathology of Alzheimer's disease". Neuron (1991), 6(4), 487-498.
[25] M. Omura, N. Sonehara and T. Okumura, "Practical approach for disease similarity calculation based on disease phenotype, etiology, and locational clues in disease names," 2016 IEEE International Conference on Bioinformatics and Biomedicine (BIBM), Shenzhen, China, 2016, pp. 1002-1009. doi:10.1109/BIBM.2016.7822659

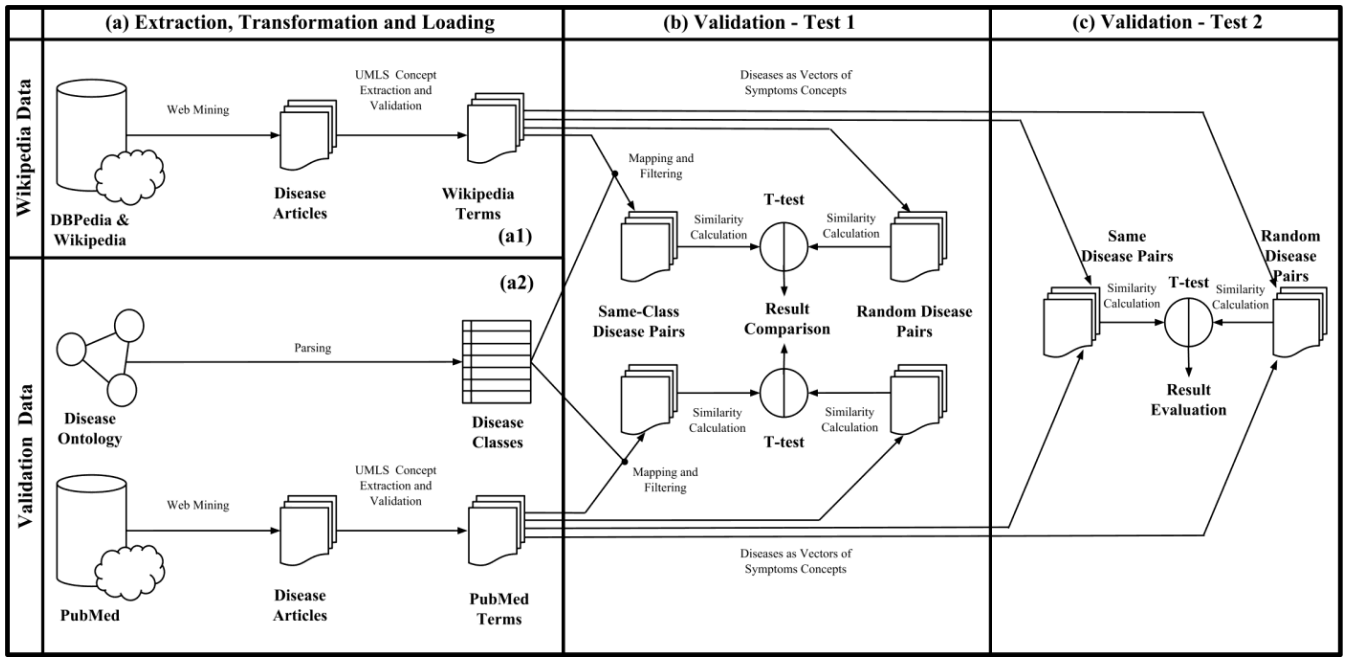

Figure 5. Schema representing the methods followed in the research, including: (a) The extraction, transformation and loading phase, in which the disease related articles were mined from Wikipedia (a1) to later obtain and validate the UMLS concepts contained on them. In order to retrieve data for the validation phase, a similar process was carried out for disease related articles from PubMed (a2), and the Disease Ontology was parsed to obtain a list of diseases by class. (b) The first validation test, in which the statistic significance of the similarity of disease pairs on a same class DO was evaluated against a set of random disease pairs. And (c), the second validation test, in which the statistic significance of the similarity of pairs of the same disease obtained from Wikipedia and PubMed was evaluated against a set of random disease pairs.